\documentstyle{article}
\advance\voffset by -2cm

\def\a{\alpha}
\def\b{\beta}

\def\g{\gamma}
\def\h{\eta}

\def\q{\theta}

\def\G{\Gamma}

\def\inbar{\vrule height1.5ex width.4pt depth0pt}
\def\rlx{\relax\leavevmode}
\def\I{\leavevmode\hbox{\small1\kern-3.8pt\normalsize1}}
\def\openone{\leavevmode\hbox{\small1\kern-3.3pt\normalsize1}}
\def\Ione{\rlx{\rm 1\kern-2.7pt l}}
\def\Ik{\rlx{\rm I\kern-.18em k}}
\def\IC{\rlx\leavevmode
             \ifmmode\mathchoice
                    {\hbox{\kern.33em\inbar\kern-.3em{\rm C}}}
                    {\hbox{\kern.33em\inbar\kern-.3em{\rm C}}}
                    {\hbox{\kern.28em\sinbar\kern-.25em{\rm C}}}
                    {\hbox{\kern.25em\ssinbar\kern-.22em{\rm C}}}
             \else{\hbox{\kern.3em\inbar\kern-.3em{\rm C}}}\fi}
\def\IP{\rlx{\rm I\kern-.18em P}}
\def\IR{\rlx{\rm I\kern-.18em R}}
\def\IN{\rlx{\rm I\kern-.20em N}}

\def\llsymbol#1{\@llsymbol{\@nameuse{c@#1}}}
\def\@llsymbol#1{\ifcase#1\or {}\or {'}\or {''}\or {'''}\or
   {''''}\or {'''''}\or  \else\@ctrerr\fi\relax}
\newcounter{contador}

\newcommand{\ol}\overline
\newcommand{\ti}\tilde
\newcommand{\wt}\widetilde
\newcommand{\wh}\widehat
\newcommand{\bv}\breve
\newcommand{\dg}\dagger

\newcommand{\aand}{\;\;\;\mbox{and}\;\;\;}

\newcommand{\be}{\begin{equation}}
\newcommand{\ee}{\end{equation}}
\newcommand{\bl}{\begin{eqnarray}&}
\newcommand{\el}{&\end{eqnarray}}
\newcommand{\bq}{\begin{eqnarray}}
\newcommand{\eq}{\end{eqnarray}}

\newcommand{\ov}{\overline}

\begin{document}

{\hfill 
\parbox{45mm}{{\large hep-th/9908190\\
                             TUW-99-19}} \vspace{3mm} }

\begin{center}
{\LARGE {\bf Superpropagators for explicitly broken $3D$-supersymmetric
theories}}

\vspace{6mm}

{\large 
J.L. Boldo$^{{\rm (a),}}$\footnote{
Supported by the {\it Conselho Nacional de Desenvolvimento
Cient\'{\i}fico e Tecnol\'{o}gico (CNPq)}.}, L.P. Colatto$^{{\rm (b)}}$, 
M.A. De Andrade$^{{\rm (a),(c)}}$, \\ O.M. Del Cima$^{{\rm (d),}}$\footnote{
Supported by the {\it Fonds zur F\"{o}rderung der
Wissen\-schaftlichen Forschung (FWF)} under the contract number
P11654-PHY.} and J.A. Helay\"{e}l-Neto$^{{\rm (a),(c),}}$\footnote{Supported in part 
by the {\it Conselho Nacional de Desenvolvimento
Cient\'{\i}fico e Tecnol\'{o}gico (CNPq)}.} }

\vspace{4mm}
$^{{\rm (a)}}${\it Centro Brasileiro de Pesquisas F\'{\i}sicas (CBPF), 
\\Departamento de Teoria de Campos e Part\'{\i}culas (DCP), \\Rua
Dr. Xavier Sigaud 150 - 22290-180 - Rio de Janeiro - RJ - Brazil.}
\vspace{2mm}

$^{{\rm (b)}}${\it Universidade de Bras\'{\i}lia (UnB),\\
Instituto de F\'{\i}sica (IF) - N\'{u}cleo de Relatividade e Teoria de
Part\'{\i}culas (NRTP),\\
Campus Darcy Ribeiro - 70790-900 - Bras\'{\i}lia - DF- Brazil.}
\vspace{2mm}

$^{{\rm (c)}}${\it Universidade Cat\'{o}lica de Petr\'{o}polis (UCP), \\ 
Grupo de F\'{\i}sica Te\'{o}rica, \\Rua Bar\~{a}o do Amazonas, 124 -
25685-070 - Petr\'{o}polis - RJ - Brazil.}\vspace{2mm}

$^{{\rm (d)}}${\it Institut f\"{u}r Theoretische Physik (ITP), \\
Technische Universit\"{a}t Wien (TU-Wien), \\Wiedner Hauptstra{\ss}e
8-10 - A-1040 - Vienna - Austria.}\vspace{4mm}


{\tt {E-mails:jboldo@cbpf.br,colatto@fis.unb.br,
marco@cbpf.br,delcima@tph73.tuwien.ac.at,helayel@cbpf.br.}}
\end{center}


\begin{abstract}
A systematic algorithm to derive superpropagators in the case of 
either explicitly or spontaneously broken supersymmetric three-dimensional 
theories is presented. We discuss how the explicit breaking terms that 
are introduced at tree-level induce 1-loop radiative corrections to 
the effective action. We also point out that the renormalisation effects 
and the breaking-inducing-breaking mechanism become more immediate 
whenever we adopt the shifted superpropagators discussed in this letter. 
\end{abstract}

Supergraph techniques have shown their efficacy since the early works on
superfield perturbation theory introduced by Salam and Strathdee 
\cite{Salam}. They also appeared to be an essential tool for the proof 
of the finiteness of the $N$$=$$2$ and $N$$=$$4$ super-Yang-Mills 
theories to all 
orders in perturbation theory \cite{Caswell, Grisaru1}. Power-counting, 
the analysis of the ultraviolet behaviour of globally supersymmetric 
and supergravity theories, and loop computations by means of super-Feynman 
rules, are much more compact and have been employed in a number of works 
to detect at which order in the perturbative series the S-matrix may 
indicate the appearance of divergences \cite{Gates1}. 

Supersymmetry however is not an exact symmetry of the low-energy world. The
breaking, either spontaneous or explicit, must be thoroughly studied not
only for phenomenological purposes, but also to check till which extent
deviations from exact supersymmetry may still be compatible with the taming
of the divergences imposed by such a symmetry. Along this line of thought,
Girardello and Grisaru \cite{Grisaru2} have sorted out a detailed
classification of all soft and hard breakings of supersymmetry in four
dimensions; their work triggered a whole line of investigation on the issue
of explicit breaking of global supersymmetry \cite{Jose}. Explicitly broken two-dimensional supersymmetric 
gauge models have been widely studied in \cite{Jose1}. In three 
dimensions, by following similar strategy of \cite{Grisaru2}, Gates and 
Nishino have been classified the soft breakings terms of $N$$=$$2$ 
supersymmetry \cite{Gates2}. The issue of partial spontaneous supersymmetry 
breaking 
$N$$=$$2$$\rightarrow$$N$$=$$1$ in $D$$=$$3$ has been studied in \cite{Zupnik}.

Recently, supersymmetry in three dimensions has been reconsidered in
connection with Yang-Mills-Chern-Simons gauge theories, which display
remarkable features as long as their ultraviolet properties are concerned, 
namely their finiteness at all orders in perturbation theory \cite{YMCS}. 
Also, with the raising of interest on supermembranes, three-dimensional 
supersymmetry becomes a major field of investigation \cite{Branes}.

Our letter sets out to reassess superfield Feynman rules whenever
supersymmetry is broken (spontaneously or explicitly) in three dimensions.
Indeed, spontaneous breakdown may always be rephrased as explicit breakings,
with explicit $\theta$-dependence, after superfields are shifted by their
vacuum expectation values.

So, for the sake of setting a systematic procedure to derive
superpropagators in the case of broken supersymmetry, we concentrate on the
explicit breakings since they naturally account for the case of spontaneous
breaking. With the results we shall present in the sequel, the reassessment
of supergraph calculations for $3D$ broken supersymmetric models becomes more
systematic and approximations introduced by simply treating the breakings as
insertions are by-passed, since we are able to sum up the latter to all
orders and so modify the superpropagators with all powers in the breaking
parameters, rather than viewing the breakings as new vertices that correct
the exact superpropagators.

We consider an explicitly broken supersymmetric theory of a complex scalar
superfield, $\Phi$, minimally coupled to a real spinor gauge superfield, 
$\Gamma_\alpha$.

In three dimensions, the most general complex scalar superfield may be 
$\theta$-expanded in component fields as follows
\footnote{The notations and conventions adopted throughout this work are 
those of ref.\cite{Gates1}, and the superspace measure adopted is 
$dv=d^3xd^2\theta$. The representation for the $\gamma$-matrices is taken 
as $\gamma^a=(\sigma_y,i\sigma_z,i\sigma_x)$, where 
$\gamma^a\equiv(\gamma^a)_\alpha^{~\beta }$ and $\{\g^a,\g^b\}=-2\h^{ab}$, 
with the metric being given by $\h^{ab}={\rm diag}(-++)$.}: 
\begin{equation}
\Phi(x,\theta)=A(x) + \theta^\alpha\psi_\alpha(x) - \theta^2F(x)~~,\label{1}
\end{equation}
with $A$ and $\psi_\alpha$ being respectively complex scalar and
two-component spinor fields, while $F$ is a complex auxiliary scalar field.
On the other hand, a three-dimensional supersymmetric gauge field theory may 
be described by a real spinor supermultiplet, 
\begin{equation}
\Gamma_\alpha(x,\theta)=\chi_\alpha(x) - \theta^\gamma[C_{\gamma\alpha}B(x) 
- iV_{\gamma\alpha}(x)] - \theta^2[2\lambda_\alpha(x) 
- i\partial_{~\alpha}^\gamma\chi_\gamma(x)]~~,\label{2}
\end{equation}
where $\chi^\alpha$ and $\lambda^\alpha$ are (real) Majorana spinors, $B$
is a real scalar, whereas $V_\alpha^{~\beta}=(\gamma^a)_\alpha^{~\beta}V_a$ 
is the gauge potential and $V_a$ being the gauge field. Also, we define the 
gauge-invariant field-strength superfield:
\begin{equation}
W_\alpha={\frac 12}D^\beta D_\alpha \Gamma_\beta~~,\label{3}
\end{equation}
constrained by $D^\alpha W_\alpha=0$.

The minimal coupling between matter and gauge superfields is accomplished by
means of the covariant supersymmetric gauge derivative:
\begin{equation}
\nabla_\alpha\Phi=D_\alpha\Phi - ig\Gamma_\alpha\Phi \aand 
\nabla_\alpha\overline{\Phi}=D_\alpha\overline{\Phi} 
+ ig\Gamma_\alpha\overline{\Phi}~~,\label{4}
\end{equation}
where $g$ is the coupling constant.

Then, we start from an action describing the broken theory of a complex
matter superfield minimally coupled to the gauge superfield in three
dimensions: 
\begin{equation}
S=S_{\rm m} + S_{\rm g}~~,\label{5}
\end{equation}
with 
\begin{equation}
S_{\rm m}=\int dv\left\{-{\frac 12}(1+2\theta^2 m_\psi)
(\nabla^\alpha\overline{\Phi})(\nabla_\alpha\Phi) + (m+\theta^2 m_A^2) 
\overline{\Phi}\Phi\right\}~~,\label{6}
\end{equation}
and 
\begin{equation}
S_{\rm g}=\frac 12\int dv\left\{(1-2\theta^2 m_\lambda) 
W^\alpha W_\alpha + \mu\Gamma^\alpha W_\alpha\right\}~~,\label{7}
\end{equation}
where $m$ and $\mu$ are the mass parameters, whereas $m_\psi$, $m_A^2$ and 
$m_\lambda$ are the coefficients for the broken terms in the matter and
gauge sectors (an explicit breaking associated to the Chern-Simons term,
namely $\theta^2\Gamma^\alpha W_\alpha$, has not been considered for such
a term also explicitly breaks gauge invariance). Besides the matter broken 
terms, $S_m$ contains kinetic and massive terms for the matter superfield 
along with the minimal coupling to the gauge superfield, while $S_g$ contains 
kinetic and the topological gauge-invariant mass terms for the gauge 
superfield.

Furthermore, the action (\ref{5}) is invariant under the following gauge
transformations:
\begin{equation}
\delta\Phi=iK\Phi\aand\delta\Gamma_\alpha={\frac 1g}D_\alpha K~~,  \label{8}
\end{equation}
where $K=K(x,\theta)$ is a real scalar superfield. In order to
obtain the superpropagators, one has to fix this gauge invariance; we add
the following gauge-fixing term to the action of eq.(\ref{5}): 
\begin{equation}
S_{\rm gf}=-{\frac 1{4\alpha}}\int dv~(D^\alpha\Gamma_\alpha)
D^2 (D^\beta\Gamma_\beta)~~.\label{9}
\end{equation}

Now we turn to the attainment of the superpropagators for the matter and
gauge sectors by taking the inverse of the wave operators
\footnote{Products of the type $XY=Z$, shall always be assumed to be 
contracted as $X^{\alpha\gamma}Y_\gamma^{~\beta}=Z^{\alpha\beta}$.}.

The bilinear piece that stems from the action for the matter superfields is
the following: 
\begin{eqnarray}
S_{\rm m}^0\!\!\!\!&=&\!\!\!\!\int dv~\biggl\{-{\frac 12}
(D^\alpha\overline{\Phi})(D_\alpha\Phi) + m \overline{\Phi}\Phi 
- \theta^2{m_\psi}(D^\alpha\overline{\Phi})(D_\alpha\Phi) 
+ \theta^2 m_A\overline{\Phi}\Phi\biggr\}\label{10} \\
\!\!\!\!&=&\!\!\!\!\int dv~\overline{\Phi}{\cal K}\Phi~~,\nonumber
\end{eqnarray}
where the operator ${\cal K}$ reads as below: 
\begin{equation}
{\cal K}=D^2 + m + m_\psi(2\theta^2D^2+\theta^\alpha D_\alpha) 
+ m_A^2\theta^2~~.\label{11}
\end{equation}

In order to invert the above wave operator and consequently obtain the
superpropagator, we shall use the projection operator formalism. The
operators associated to the scalar superfield are classified as follows:
\begin{equation}
P_1=D^2~~,~~~P_2=\theta^2~~,~~~P_3=\theta^\alpha D_\alpha~~,~~~
P_4=\theta^2D^2 \aand P_5=i\partial_{\alpha\beta}\theta^\alpha D^\beta~~,
\label{12}
\end{equation}
and their operator algebra is displayed in Table \ref{table1}. Moreover, 
we present some useful relations: 
\begin{eqnarray}
&&\left\{D_\alpha,\theta_\beta\right\}=C_{\alpha\beta}~~,~~~
\left\{D_\alpha,\theta^\beta\right\}=\delta_\alpha^{~\beta}~~,~~~
\left\{D^\alpha,\theta_\beta\right\}=-\delta_\beta^{~\alpha}~~,~~~
\left\{D^\alpha,\theta^\beta\right\}=C^{\alpha\beta}~~,\label{13} \nonumber\\
&&[D^2,\theta_\alpha]=D_\alpha~~,~~~[D_\alpha,\theta^2]=\theta_\alpha 
\aand [D^2,\theta^2]=-1+\theta^\alpha D_\alpha~~. 
\end{eqnarray}

Thus, rewriting ${\cal K}$ in terms of the operators $P_{\hat \i}$ (${\hat \i}=0,1,...,5$), we have 
\begin{equation}
{\cal K}=mP_0 + P_1 + m_A^2P_2+{m_\psi}P_3 + 2m_\psi P_4~~,\label{14}
\end{equation}
where $P_0\equiv1$.
\begin{table}
\centering
\begin{tabular}{|c|c|c|c|c|c|}
\hline
& $P_1$ & $P_2$ & $\,\,\,\,\,\,\,\,P_3$ & $P_4$ & $P_5$ \\ \hline
$P_1$ & $\Box $ & $-1+P_3+P_4$ & $2P_1+P_5$ & $-P_1+\Box P_2-P_5$ & $\Box
(-2+P_3)$ \\ \hline
$P_2$ & $P_4$ & $0$ & $0$ & $0$ & $0$ \\ \hline
$P_3$ & $-P_5$ & $2P_2$ & $P_3-2P_4$ & $2P_4$ & $2\Box P_2+P_5$ \\ \hline
$P_4$ & $\Box P_2$ & $-P_2$ & $2P_4$ & $-P_4$ & $-2\Box P_2$ \\ \hline
$P_5$ & $-\Box P_3$ & $0$ & $-2\Box P_2+P_5$ & $\,0$ & $\Box \left(
P_3+2P_4\right) $ \\ \hline
\end{tabular}
\caption[t1]{Multiplicative table fulfilled by $P_1$, $P_2$, $P_3$, $P_4$ and 
$P_5$. The products are supposed to be in the ordering ``row times column''.}
\label{table1}
\end{table}

Using the algebra of Table \ref{table1}, we readily obtain the 
superpropagator: 
\begin{equation}
\left\langle 0\right|T[\overline{\Phi}(x_1,\theta_1)
\Phi(x_2,\theta_2)]\left|0\right\rangle=i{\cal K}_{\theta_1}^{-1}\delta^3
(x_1-x_2)\delta(\theta_1-\theta_2)~~,\label{15}
\end{equation}
where we are using 
\begin{equation}
\delta^3(x_1-x_2)=\int\frac{d^3k}{(2\pi)^3}~e^{-ik(x_1-x_2)}~~.\label{16}
\end{equation}
In momentum space, (\ref{15}) is given by: 
\begin{eqnarray}
\left\langle\overline{\Phi}(k,\q_1)\Phi(k,\q_2)\right\rangle\!\!\!\!&=&\!\!\!\!
\frac{-i}{k^2+m^2+m_A^2}\biggl\{mP_0-P_1 + \nonumber\\
&&-~\frac 1{k^2+(m+m_\psi)^2}\biggl[[k^2(m_\psi^2+mm_\psi) 
+ m_A^2(m+m_\psi)^2-m_A^2k^2]P_2 + \nonumber\\
&&+~[k^2m_\psi +m_A^2(m+m_\psi) - m(m_\psi^2-mm_\psi)](P_3+2P_4) + 
\nonumber\\
&&+~(m_A^2-m_\psi^2-mm_\psi) P_5\biggr]\biggr\}
\delta(\theta_1-\theta_2)~~.
\end{eqnarray}
From the poles in $k^2$, it becomes clear that the physical scalar and the
fermion field have their masses shifted with respect to the degenerated
value corresponding to exact supersymmetry.

The propagators for the component fields can be read off by making use of
the following relations: 
\begin{eqnarray}
\delta(\theta_1-\theta_2)\!\!\!\!&=&\!\!\!\!-(\theta_1-\theta_2)^2~~,
\nonumber \\
\theta_1^2\delta(\theta_1-\theta_2)\!\!\!\!&=&\!\!\!\!-\theta_1^2\theta_2^2~~,
\nonumber \\
\theta_1^\alpha D_{1\alpha}\delta(\theta_1-\theta_2)\!\!\!\!&=&\!\!\!\!
-2\theta_1^2 + \theta_1\theta_2~~,\nonumber \\
\theta_1^2D_1^2\delta(\theta_1-\theta_2)\!\!\!\!&=&\!\!\!\!\theta_1^2~~,
\nonumber \\
k_{\alpha\beta}\theta_1^\alpha D_1^\beta\delta(\theta_1-\theta_2) 
\!\!\!\!&=&\!\!\!\!k_{\alpha\beta}\theta_1^\alpha\theta_2^\beta 
+ 2k^2\theta_1^2\theta_2^2~~.\label{18}
\end{eqnarray}
The non-vanishing component-field propagators may be extracted out of 
$\left\langle\overline{\Phi}(k,\q_1)\Phi(k,\q_2)\right\rangle$: 
\begin{eqnarray}
\left\langle{\ov A}(k)A(k)\right\rangle\!\!\!\!&=&\!\!\!\!
\frac{-i}{k^2+m^2+m_A^2}~~, 
\nonumber \\ 
\left\langle {\ov F}(k)F(k)\right\rangle\!\!\!\!&=&\!\!\!\!i\frac{k^2+m_A^2}
{k^2+m^2+m_A^2}~~,\nonumber \\
\left\langle {\ov A}(k)F(k)\right\rangle\!\!\!\!&=&\!\!\!\!
\frac{im}{k^2+m^2+m_A^2}~~,
\nonumber \\
\left\langle\ov\psi^\alpha(k)\psi^\beta(k)\right\rangle\!\!\!\!&=&\!\!\!\!
-i\frac{k^{\alpha\beta}-(m+m_\psi)C^{\alpha\beta}}{k^2+(m+m_\psi)^2}~~,
\label{19}
\end{eqnarray}
which agree with the propagators calculated from the component-field action
stemming from eq.(\ref{6}).

As for the gauge sector, similarly to the matter sector, we may find the
superpropagator for the gauge superfield. The bilinear piece of (\ref{5})
for the gauge superfield plus the gauge-fixing term, (\ref{9}), reads 
\begin{eqnarray}
S_{\rm g}^0\!\!\!\!&=&\!\!\!\!
\frac 12\int dv~\biggl\{{\frac 14}(D^\beta D^\alpha\Gamma_\beta)
(D^\gamma D_\alpha\Gamma_\gamma) + {\frac \mu 2}\Gamma^\alpha
(D^\beta D_\alpha\Gamma_\beta) 
- {\frac 1{2\alpha}}(D^\alpha\Gamma_\alpha)(D^2D^\beta\Gamma_\beta) + 
\nonumber \\
&&-~\theta^2\frac{m_\lambda }2(D^\beta D^\alpha\Gamma_\beta)
(D^\gamma D_\alpha\Gamma_\gamma )\biggr\}  \nonumber \\
\!\!\!\!&=&\!\!\!\!\int dv~\Gamma_\alpha{\cal K}^{\alpha\beta}\Gamma_\beta~~, 
\label{20}
\end{eqnarray}
where the operator ${\cal K}^{\alpha\beta}$ can be written as 
\begin{equation}
{\cal K}^{\alpha\beta}=-{\frac 14}\left[{\frac 12}D^\gamma D^\alpha
D^\beta D_\gamma + {\frac 1\alpha}D^\alpha D^2D^\beta + \mu D^\beta D^\alpha
- m_\lambda D^\gamma D^\alpha\theta^2D^\beta D_\gamma\right]~~.  \label{21}
\end{equation}

Here, we must introduce other twelve superspace operators coming from the
gauge sector, which can be expressed in terms of the $P_{\hat \i}$'s as follows:
\begin{equation}
R_{\hat \i}^{\alpha\beta}=iP_{\hat \i}\partial^{\alpha\beta}\aand
S_{\hat \i}^{\alpha\beta}=P_{\hat \i}C^{\alpha\beta}~~,\label{22}
\end{equation}
where ${\hat \i}=0,1,...,5$. Their algebra is presented in Table \ref{table2}.
\begin{table}
\centering
\begin{tabular}{|c|c|c|c|c|}
\hline
& $S_0$ & $S_{\rm j}$ & $R_0$ & $R_{\rm j}$ \\ \hline
$S_0$ & $S_0$ & $S_{\rm j}$ & $R_0$ & $R_{\rm j}$ \\ \hline
$S_{\rm i}$ & $S_{\rm i}$ & $P_{\rm i}P_{\rm j}C^{\alpha\beta}$ & $R_{\rm i}$ & $i\Box P_{\rm i}
P_{\rm j}\partial^{\alpha\beta}$ \\ \hline
$R_0$ & $R_0$ & $R_{\rm j}$ & $\Box S_0$ & $\Box S_{\rm j}$ \\ \hline
$R_{\rm i}$ & $R_{\rm i}$ & $i\Box P_{\rm i}P_{\rm j}\partial^{\alpha\beta}$ & $\Box S_{\rm i}$ & 
$\Box P_{\rm i}P_{\rm j}C^{\alpha\beta}$ \\ \hline
\end{tabular}
\caption[t2]{Multiplicative table fulfilled by $R_{\hat \i}^{\alpha\beta}$ 
and $S_{\hat \i}^{\alpha\beta}$. The ordering is ``row times column''.}
\label{table2}
\end{table}

By using the property $P_{\rm i}P_{\rm j}=\sum a_{\rm k} P_{\rm k}$ (see Table \ref{table1}) 
and (\ref{22}), where ${\rm i},{\rm j},{\rm k}=1,2,...,5$, we may check that the algebra 
presented in Table \ref{table2} is in fact closed. Thus, the wave 
operator ${\cal K}^{\alpha\beta}$ can be rewritten as
\begin{eqnarray}
{\cal K}^{\alpha\beta}\!\!\!\!&=&\!\!\!\!
-\frac 14\biggl[\frac{\alpha+1}\alpha\Box S_0^{\alpha\beta} 
+ (\mu+m_\lambda)S_1^{\alpha\beta} - 2m_\lambda\Box S_2^{\alpha\beta} 
+ m_\lambda S_5^{\alpha\beta} + \nonumber\\
&&+~(\mu+m_\lambda)R_0^{\alpha\beta} 
+ \frac{\alpha -1}\alpha R_1^{\alpha\beta} - m_\lambda R_3^{\alpha\beta} 
- 2m_\lambda R_4^{\alpha\beta}\biggr]~~.\label{23} 
\end{eqnarray}

The superpropagator 
\begin{equation}
\left\langle 0\right|T[\G^\a(x_1,\theta_1)
\G^\b(x_2,\theta_2)]\left|0\right\rangle=i{\cal K}_{\theta_1}^{-1\a\b}\delta^3
(x_1-x_2)\delta(\theta_1-\theta_2)~~,\label{propg}
\end{equation}
exhibits the following structure in terms of the
superspace operators: 
\begin{equation}
\left\langle 0\right|T[\G^\a(x_1,\theta_1)
\G^\b(x_2,\theta_2)]\left|0\right\rangle=
\sum_{\hat \i=0}^5(s_{\hat \i}S^{\a\b}_{\hat \i} + r_{\hat \i}R^{\a\b}_{\hat \i})
\delta^3(x_1-x_2)\delta(\theta_1-\theta_2)~~,
\label{24}
\end{equation}
where the coefficients $s_{\hat \i}$ and $r_{\hat \i}$ are to be (uniquely) 
determined by a system of twelve equations. The latter are solved and the 
answer we find, in momentum space, is presented below: 
\begin{eqnarray}
\left\langle\Gamma^\alpha(k,\q_1)\Gamma^\beta(-k,\q_2)\right\rangle
\!\!\!\!&=&\!\!\!\!\frac{-i}{k^2+(\mu+m_\lambda)^2}\biggl\{(\alpha+1)
S_0^{\alpha\beta} + \frac{\alpha-1}{k^2}R_1^{\alpha\beta} + \nonumber \\
&&+~\frac{\mu+m_\lambda}{k^2}\left[\alpha(\mu+m_\lambda)
\left(S_0^{\alpha\beta}+\frac 1{k^2}R_1^{\alpha\beta}\right) 
+ S_1^{\alpha\beta} + R_0^{\alpha\beta}\right] +  \nonumber \\
&&+~\frac{m_\lambda(2\mu+m_\lambda)}{k^2+\mu^2}
\left(S_3^{\alpha\beta} + 2S_4^{\alpha\beta} + 2R_2^{\alpha\beta} 
+ \frac 1{k^2}R_5^{\alpha\beta}\right) +  \nonumber \\
&&+~\frac{m_\lambda[k^2-\mu(\mu+m_\lambda)]}{k^2(k^2+\mu^2)}
\left(2k^2S_2^{\alpha\beta} + S_5^{\alpha\beta} - R_3^{\alpha\beta} 
- 2R_4^{\alpha\beta}\right)\biggr\}\delta(\theta_1-\theta_2)~~. \label{25} 
\end{eqnarray}

Again, two simple poles show up which correspond to the gauge boson and the
gaugino masses; as shown below, they are split according to:
\begin{eqnarray}
\left\langle\chi^\alpha(k)\chi^\beta(-k)\right\rangle\!\!\!\!&=&\!\!\!\!
-i\frac{[\alpha(\mu+m_\lambda)^2 + (\alpha -1)k^2]k^{\alpha\beta} 
+ (\mu+m_\lambda)k^2C^{\alpha\beta}}
{k^4[k^2+(\mu+m_\lambda)^2]}~~,\nonumber \\
\left\langle\chi^\alpha(k)\lambda^\beta(-k)\right\rangle\!\!\!\!&=&\!\!\!\!
-i\frac{(\mu+m_\lambda)k^{\alpha\beta} + k^2C^{\alpha\beta}}
{k^2[k^2+(\mu+m_\lambda)^2]}~~,\nonumber \\
\left\langle\lambda^\alpha(k)\lambda^\beta(-k)\right\rangle\!\!\!\!&=&\!\!\!\!
-i\frac{k^{\alpha\beta} - (\mu+m_\lambda)C^{\alpha\beta}}
{k^2+(\mu+m_\lambda)^2}~~,\nonumber \\
\left\langle B(k)B(-k)\right\rangle\!\!\!\!&=&\!\!\!\!
i\frac{\alpha}{k^2}~~,\nonumber \\
\left\langle V^a(k)V^b(-k)\right\rangle\!\!\!\!&=&\!\!\!\!
\frac{-i}{k^2+\mu^2}\left(\eta^{ab} - \frac{k^ak^b}{k^2}\right) 
- i\alpha \frac{k^ak^b}{k^4} 
- \frac \mu {k^2(k^2+\mu^2)}\varepsilon^{abc}k_c~~.\label{26}
\end{eqnarray}

Once we worked out the matter and gauge superpropagators with the breaking
parameters summed up to all orders, we have at our disposal enough data to
discuss how the explicit breakings that are introduced at tree-level induce
1-loop radiative corrections to the effective action. We adopt the
superfield Feynman rules as presented and discussed in ref.\cite{Gates1} and
make use of the superpropagators derived in our paper.

If there are matter superfields in sufficient number such that 
$\Phi^3$-interaction vertices do not break gauge invariance, tadpole 
supergraphs with a $\Phi$-superfield on the external leg may induce a 
loop correction to the $F$-term, as a result of the term with the operator 
$P_4$ present in the $\overline{\Phi}\Phi$-propagator. $F$-terms are not 
radiatively induced as a result of the gauge interaction, since the gauge 
couplings do not allow a $\Phi$-tadpole with a loop where the gauge 
superfield flow inside alone. It is the matter self-interaction and the 
explicit breakings governed by $m_\psi$ and $m_A$ the responsible for 
the 1-loop generation of an $F$-term.

Also, it is interesting to notice that, if we consider the 2-point function
with $\Gamma_{\alpha}$ and $\Gamma_{\beta}$ on the external legs and the
matter superpropagators running inside the loop, the matter breaking terms
induce a 1-loop correction to the supersymmetric Chern-Simons mass term; as
for the gauge superfield kinetic term, no correction arises that comes from
the breakings. The technical reason to understand these results is a simple
counting of covariant derivatives inside the loop (some are brought by the
propagators, others appear as vertex factors).

The gauge-invariant term that splits the scalar mass inside the matter
supermultiplet ($\theta^2\overline{\Phi}\Phi$) induces a 1-loop
correction to the breaking term that splits the gaugino mass in the gauge
supermultiplet ($\theta^2W^\alpha W_\alpha$): such a term appears as the
result of the interference between breaking terms present in the matter
superpropagators. Conversely, it is noteworthy to remark that the gaugino
mass breaking term yields a 1-loop correction that explicitly breaks
supersymmetry and splits the scalar mass inside the matter multiplet. The
latter result can be readily attained if we compute a 1-loop diagram with
matter superfields on the external legs and a gauge superpropagator
appearing as an internal line of the corresponding 1-loop graph.

Concluding these comments, one of the advantages of working with this
somewhat complicated superpropagators is that, once the breaking parameters
are taken into account to all orders, we get a safe and systematic algorithm
for deriving radiative corrections to the breakings as induced from one
another. The investigation of the effective action, renormalisation effects
and breaking-inducing-breaking mechanism become more automatic if we adopt
to work with these full superpropagators. In situations where a
spontaneous breaking of supersymmetry takes place, and shifts have to be
performed around the true ground state, explicit breakings as the ones
collected above show up (spontaneous supersymmetry breaking appears 
in superspace as $\theta$-terms) and our computations may become useful 
to compute radiative corrections to the effective action and to physical 
quantities derived from the effective potential. In the case supersymmetry 
is broken for a gauge model, we have to generalize the $R_\xi$-gauge with 
now $\theta$-dependent present, since the superfields $\Phi$ and 
$\Gamma_\alpha$ mix up with a $\theta^2$-factor. This problem is now under 
investigation, and we shall soon report our results elsewhere.

\underline{Acknowledgements}:
Two of us (L.P.C.) and (O.M.D.C.) thank {\it The Erwin Schr\"odinger 
International Institute for Mathematical Physics (ESI-Vienna)} for 
the kind hospitality, they also would like to express their gratitude 
to Manfred Schweda and Wolfgang Kummer. (L.P.C.) thanks also the 
{\it Institut f\"ur Theoretische Physik} of the {\it TU-Wien}. 
(O.M.D.C.) acknowledges the {\it Departamento de Teoria de Campos e 
Part\'\i culas (DCP)} of the {\it Centro Brasileiro de Pesquisas 
F\'\i sicas (CBPF)} for the warm hospitality during all his visits. 
He dedicates this work to his wife, Zilda Cristina, to his kids, Vittoria 
and Enzo, and his mother, Victoria.


\begin{thebibliography}{9}
\bibitem{Salam} A. Salam and J. Strathdee, Nucl.Phys. B86 (1975) 142 and
Fortsch.Phys. 26 (1978) 57.

\bibitem{Caswell} W.E. Caswell and D. Zanon, Phys.Lett. B100 (1981) 152 
and Nucl.Phys. B182 (1981) 125.

\bibitem{Grisaru1} M.T. Grisaru, M. Ro\v{c}ek and W. Siegel, Nucl.Phys. B183 
(1981) 141.

\bibitem{Gates1} S.J. Gates Jr., M.T. Grisaru, M. Ro\v{c}ek and W. Siegel,
{\it ``Superspace: or One Thousand and One Lessons in Supersymmetry''},
Benjamin/Cummings (London-UK), 1983.

\bibitem{Grisaru2} L. Girardello and M.T. Grisaru, Nucl.Phys. B194 (1982) 65.

\bibitem{Jose} J.A. Helay\"{e}l-Neto, Phys.Lett. B135 (1984) 78; 
F. Feruglio, J.A. Helay\"{e}l-Neto, F. Legovini, Nucl.Phys. B249 (1985) 533;
J.A. Helay\"{e}l-Neto, F.A.B.R. de Carvalho and A.W. Smith, Nucl.Phys. 
B271 (1986) 175 and Nucl.Phys. B278 (1986) 309.

\bibitem{Jose1} N. Chair, J.A. Helay\"{e}l-Neto and 
A.W. Smith, Phys.Lett. B195 (1987) 407; J.A. Helay\"{e}l-Neto and A.W. Smith, Phys.Lett. B196 (1987) 503 and 507; and 
Int.J.Mod.Phys. A9 (1990) 1861.

\bibitem{Gates2} S.J. Gates Jr. and H. Nishino, Phys.Lett. B281 (1992) 72.

\bibitem{Zupnik} B.M. Zupnik, {\it Partial spontaneous breakdown of 
$3$-dimensional $N$$=$$2$ supersymmetry}, hep-th/9905108.

\bibitem{YMCS} F. Ruiz Ruiz and P. van Nieuwenhuizen, Nucl.Phys. B486
(1997) 443, hep-th/9609074; O.M. Del Cima, D.H.T. Franco, 
J.A. Helay\"el-Neto and O. Piguet, J.High Energy Phys.(JHEP) 02 (1998) 002, 
hep-th/9711191, and Lett.Math.Phys. 47 (1999) 265, math-ph/9904030.

\bibitem{Branes} O. Bergman, A. Hanany, A. Karch and B. Kol, 
{\it Branes and supersymmetry breaking in $3D$ gauge theories}, 
hep-th/9908075; K. Ohta, {\it Supersymmetric index and $s$-rule for 
type IIB branes}, hep-th/9908120.
\end{thebibliography}
\end{document}